\documentclass[twocolumn,superscriptaddress,amsmath,amssymb,aps,pra,floatfix]{revtex4-1}

\usepackage[utf8]{inputenc}
\usepackage{wrapfig}
\usepackage{float}
\usepackage{amssymb}
\usepackage{amsfonts}
\usepackage{amsmath}
\usepackage{cases}
\usepackage{stmaryrd}\usepackage{tipa}
\usepackage[dvips]{graphicx}
\usepackage{dcolumn}
\usepackage{bbold}
\usepackage{graphicx}
\usepackage{graphicx}
\usepackage{subfigure}
\usepackage{dcolumn}
\usepackage{bm}
\usepackage{color}
\usepackage{xcolor}
\usepackage{CJK}
\usepackage{subfigure}
\usepackage{physics}
\usepackage{array}
\usepackage{multirow}
\usepackage{nopageno}
\usepackage{mathtools}
\usepackage{comment}
\usepackage{enumerate}
\usepackage{ragged2e}

\usepackage{array}

\begin{document}

\title{Quantized Single-Ion-Channel Hodgkin-Huxley Model for Quantum Neurons}

\author{Tasio Gonzalez-Raya}
\affiliation{Department of Physical Chemistry, University of the Basque Country UPV/EHU, Apartado 644, 48080 Bilbao, Spain}

\author{Xiao-Hang Cheng}
\affiliation{Department of Physical Chemistry, University of the Basque Country UPV/EHU, Apartado 644, 48080 Bilbao, Spain}
\affiliation{International Center of Quantum Artificial Intelligence for Science and Technology (QuArtist) \\ and Department of Physics, Shanghai University, 200444 Shanghai, China}

\author{I\~nigo L. Egusquiza}
\affiliation{Department of Theoretical Physics and History of Science, University of the Basque Country UPV/EHU, Apartado 644, 48080 Bilbao, Spain}

\author{Xi Chen}
\affiliation{Department of Physical Chemistry, University of the Basque Country UPV/EHU, Apartado 644, 48080 Bilbao, Spain}
\affiliation{International Center of Quantum Artificial Intelligence for Science and Technology (QuArtist) \\ and Department of Physics, Shanghai University, 200444 Shanghai, China}

\author{Mikel Sanz}
\email{mikel.sanz@ehu.eus}
\affiliation{Department of Physical Chemistry, University of the Basque Country UPV/EHU, Apartado 644, 48080 Bilbao, Spain}

\author{Enrique Solano}
\affiliation{Department of Physical Chemistry, University of the Basque Country UPV/EHU, Apartado 644, 48080 Bilbao, Spain}
\affiliation{International Center of Quantum Artificial Intelligence for Science and Technology (QuArtist) \\ and Department of Physics, Shanghai University, 200444 Shanghai, China}
\affiliation{IKERBASQUE, Basque Foundation for Science, Maria Diaz de Haro 3, 48013 Bilbao, Spain}

\begin{abstract}
The Hodgkin-Huxley model describes the behavior of the cell membrane in neurons, treating each part of it as an electric circuit element, namely capacitors, memristors, and voltage sources. We focus on the activation channel of potassium ions, due to its simplicity, while keeping most of the features displayed by the original model. This reduced version is essentially a classical memristor, a resistor whose resistance depends on the history of electric signals that have crossed it, coupled to a voltage source and a capacitor. Here, we will consider a quantized Hodgkin-Huxley model based on a quantum memristor formalism. We compare the behavior of the membrane voltage and the potassium channel conductance, when the circuit is subjected to AC sources, in both classical and quantum realms. Numerical simulations show an expected adaptation of the considered channel conductance depending on the signal history in all regimes. Remarkably, the computation of higher moments of the voltage manifest purely quantum features related to the circuit zero-point energy. Finally, we study the implementation of the Hodgkin-Huxley quantum memristor as an asymmetric rf SQUID in superconducting circuits. This study may allow the construction of quantum neuron networks inspired in the brain function, as well as the design of neuromorphic quantum architectures for quantum machine learning.
\end{abstract}

\maketitle
\section{Introduction}
Brain science and neurophysiology are fascinating topics posing deep questions regarding the global comprehension of the human being. Understanding how the brain works catalyzed interdisciplinary research fields such as biophysics and bioinformatics. In 1963, the Nobel Prize in Physiology or Medicine was awarded to Alan Lloyd Hodgkin and Andrew Fielding Huxley for their work describing how electric signals in neurons propagate through the axon. Their work involved modeling small segments of the axon membrane as an electric circuit represented by a set of nonlinear differential equations~\cite{HHa, HHb, HHc, HHM}, establishing a bridge between neuroscience~\cite{NeuronRev, Neuronbook} and physics~\cite{CohResHH, CohResHHNoise, NoiseHHM, ColorNoiseHH, TemporalCohHH, TimeScalesHH, IonChannelNoiseHH, SynTimeDelayHH}.

A neuron is an electrically excitable cell that receives, processes, and transmits information through electric signals, whose main components are the cell body, the dendrites, and the axon. Dendrites are ramifications that receive and transmit stimuli into the cell body, which processes the signal. The nerve impulse is then propagated through the axon, which is an extension of the nerve cell. This propagation gradient is generated through the change in the ion permeability of the cell membrane when an impulse is transmitted. This implies a variation in ion concentrations represented in the Hodgkin-Huxley circuit by a nonlinear conductance. Its resistance depends on the history of electric charges crossing the cell, which is naturally identified with a memristor~\cite{Mem, MemDev}.

In the last decade, we have witnessed a blossoming of quantum platforms for quantum technologies, where quantum simulations, quantum computing, quantum sensing, and quantum communication are to be highlighted. Superconducting circuits are one of the leading quantum platforms, allowing controllability, scalability, and coherence. The combination of biosciences with quantum technologies gave rise to a variety of novel fields, as quantum artificial intelligence~\cite{QML Lloyd}, quantum biology~\cite{QBio,  QBiobook}, quantum artificial life~\cite{QAlife,QAlife_IBM}, and quantum biomimetics~\cite{Biomim}, among others. From a wide perspective, we are heading toward a general framework for neuromorphic quantum computing, in which brain-inspired architectures strive to take advantage of quantum features to enhance computational power.

Recent proposals for a quantized memristor~\cite{QMem}, as well as its possible implementations in both superconducting circuits~\cite{CircuitQMem} and integrated quantum photonics~\cite{OptQMem}, allow the construction of a quantized neuron model based on the Hodgkin-Huxley circuit. In the classical realm this model reproduces the characteristic adaptive behavior of brain neurons, whereas, in the quantum regime it could reveal unique characteristics or an unprecedented learning performance. Furthermore, the study of memristor-based electric circuits~\cite{CoupledMem, CompositeMem, TheoremMem, AnalogMem} sets a starting point for the investigation of coupled quantum memristors, which may lead to the development of quantum neuron networks.

Classical models comprising features of real neurons are an active research field, leaving open an extension to the quantum domain~\cite{QuanDotNeuron, QubitNeuron, MirrorNeuron, ArtificialNeuron, AAGuik, LearningNeuron}. This led to the prolongation of neuromorphic computing toward the quantum realm. Neuromorphic quantum computing consists of the design of artificial structures that mimic neurobiological architectures present in the nervous system with the purpose of enhancing and accelerating certain computational tasks, utilizing quantum resources~\cite{NQC}. Furthermore, classical memristive devices have been used in the simulation of synaptic processes~\cite{Synap, SynapticDynamics, Neuromorphic} as well as learning processes~\cite{STDP}. However, none of those studies ever considered the quantized version of the Hodgkin-Huxley model.

In this article, we study a simplified version of the Hodgkin-Huxley model, preserving its biological interest, in which only the potassium-ion channel plays a role. The ion-channel conductance, modeled by a memristor, is coupled to a voltage source and a capacitor, where we study its response under periodic driving in both the classical regime and the quantum regime. By means of a quantum memristor~\cite{QMem}, we quantize the elements of the Hodgkin-Huxley circuit considered, comparing the membrane voltage, the conductance, and the \textit{I-V} curve for a classical input source~\cite{Thesis}. To conclude, we consider the implementation of the quantum memristor in the Hodgkin-Huxley circuit as an asymmetric rf superconducting quantum-interference device (SQUID) in superconducting circuits. This work establishes a road map for the experimental construction of quantum neuron networks, as well as neuromorphic quantum architectures and quantum neural networks~\cite{PercepMem}. These concepts could find general applications in the field of quantum machine learning~\cite{IntroQML} and could be a challenging alternative to a gate-based universal quantum computer.

\section{Theoretical Framework}
\subsection{Classical Hodgkin-Huxley model}

\begin{figure}[t]
{\includegraphics[width=0.5 \textwidth]{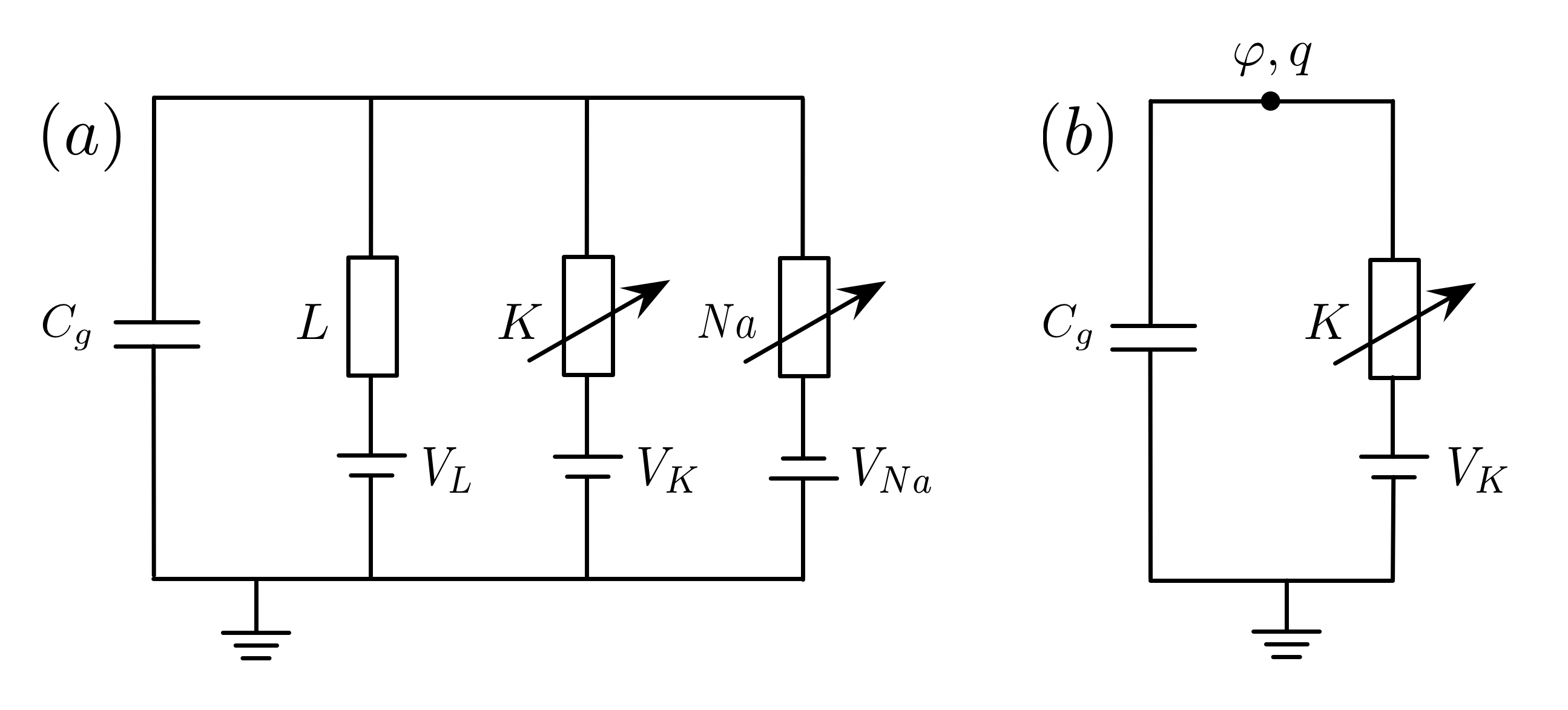}}
\caption{(a) Complete Hodgkin-Huxley circuit and (b) Hodgkin-Huxley circuit with solely a potassium-ion channel. The capacitance $C_{g}$ describes the axon's membrane capacity to store charge, the voltage sources $V_{i}$ account for the initial ion concentrations in the axon, and the ion permeabilities of the membrane are described by nonlinear conductances $i$, for $i=L,K,Na$.}
\label{circuit}
\end{figure}

The cell membrane of a neuron shows permeability changes for different ion species after receiving electric impulses through the dendrites. These changes result in variations in ion concentrations which, when a certain threshold is overcome, can lead to a sudden depolarization of the membrane and the consequent transmission of the signal through the axon. These ions comprise mainly potassium and sodium ions, which have different roles during a potential spike. In 1952, Hodgkin and Huxley developed a model that describes the propagation of these stimuli by treating each component of the excitable cell membrane as an electric circuit element, as shown in Fig.~\ref{circuit}(a). The equations of this circuit are
\begin{eqnarray}
\label{current}I(t)&=&C_g\frac{d V_g}{dt}+\bar{g}_{\text{K}} n^4 (V_g-V_{\text{K}})+\bar{g}_{\text{Na}} m^3 h (V_g-V_{\text{Na}})\nonumber\\
&&+\bar{g}_{\text{L}}(V_g-V_{\text{L}}),\\
\label{mu}\frac{dn}{dt}&=&\alpha_n(V_g)(1-n)-\beta_n(V_g)n,\\
\label{m}\frac{dm}{dt}&=&\alpha_m(V_g)(1-m)-\beta_m(V_g)m,\\
\label{h}\frac{dh}{dt}&=&\alpha_h(V_g)(1-h)-\beta_h(V_g)h,
\end{eqnarray}
where $I$ is the total current, $\bar{g}_i$ $(i=L, K, N_a)$ are constants representing the maximum value each electrical conductance can take, $C_g$ is the membrane capacitance, $V_i$ is the resting potential in each ion channel, and $n, m,$ and $h$ are dimensionless quantities between 0 and 1 that represent the probability of activation or inactivation of each ion channel. 

Except for the leakage channel \textit{L}, which accounts for small unperturbed flow of noninvolved (mostly chloride) ions, and thus is described by a constant conductance, any ion channel is characterized by a nonlinear conductance due to \textit{n, m} and \textit{h}, which weight each channel differently and depend on voltage and time. We keep solely the contribution coming from the potassium channel, which has only activation gates, and not both activation and inactivation gates as for the sodium channel. This model still conserves the most-characteristic behavior of neurons. In this case, we are left with two coupled differential equations,
\begin{eqnarray}
\label{current1}I(t)&=&C_g\frac{d V_g}{dt}+g_{\text{K}} n^4 (V_g-V_{\text{K}}),\\
\label{mu1}\frac{dn}{dt}&=&\alpha_{n}(V_g)(1-n)-\beta_{n}(V_g) n.
\end{eqnarray}
Naturally, this nonlinear conductance can be identified with a memristor.

\subsection{Memristor}
A memristor is a resistor whose resistance depends on the history of electric signals, voltages or charges, that have crossed it. This particular description of a resistor can be given in the context of Kubo's response theory~\cite{Kubo}, and it was in 1971 when Chua~\cite{Mem} rescued this idea and coined the term ``memristor''. In the original work by Hodgkin and Huxley, the memristor appears not as an independent electric circuit element but as a nonlinear resistor governed by the equations of a memristor. This identification came about years later, and is highlighted in Ref.~\cite{HHMem}. The equations describing the physical properties and memory effects of a (voltage-controlled) memristor are
\begin{eqnarray}
\label{I-V}I(t)=G(\mu(t))V(t),\\
\label{mudyn}\dot{\mu}(t)=f(\mu(t),V(t)).
\end{eqnarray}
$G(\cdot)$ and $f(\cdot)$ are continuous real functions satisfying the following conditions:
\begin{enumerate} [(i)]
\item $G(\mu) \geq 0$ for all values of $\mu$.
\item For fixed $\mu$, $f(\mu,V)$ is monotone, and $f(\mu,0)=0$.
\end{enumerate}

Property (i) implies that $G(\mu)$ can indeed be understood as a conductance, so Eq.~(\ref{I-V}) can be interpreted as a state-dependent Ohm's law. This ensures that the memristor is a passive element. Property (ii) restricts the internal variable dynamics to provide nonvanishing memory effects for all significant voltage inputs, implying that it does not have dynamics in the absence of voltage.

Solving Eq.~(\ref{mudyn}) requires time integration over the past of the control signal, and this solution affects $G(\mu)$. This means that the response in the current given by Eq.~(\ref{I-V}) depends not only on the present value of the control voltage, but also on the previous values. Hence, if a memristor is subjected to a periodic control signal, the $I-V$ curve will display a hysteresis loop, which contains memory effects, identifying the slope of this curve with the resistance of the device.

The study of a single ion channel means neutralizing the dynamics of the other channel, which is a matter of controlling its voltage; namely, its ionic concentration. Experimentally, an isolated study of the potassium channel is achieved by setting the membrane voltage to $V_{Na}$, or by changing the sodium concentration by replacing the sodic medium by any other nongated substance. This was performed in the work reported in Ref.~\cite{HHM}, and experimental results were obtained that could be compared with theoretical results, thus testing the proposed expressions for $\alpha_{x}$ and $\beta_{x}$, with $x={n, m, h}$ separately for each channel, so that values for $n$, $m$, and $h$ could be obtained. Then, with the conductance of the potassium channel given by $g_{K}=\bar{g}_{K}n^{4}$, Eqs.~\ref{current1} and~\ref{mu1} can be solved numerically for the membrane voltage, given an input current $I(t)$.

\subsection{Circuit Quantization}

\begin{figure}[]
{\includegraphics[width=0.448 \textwidth]{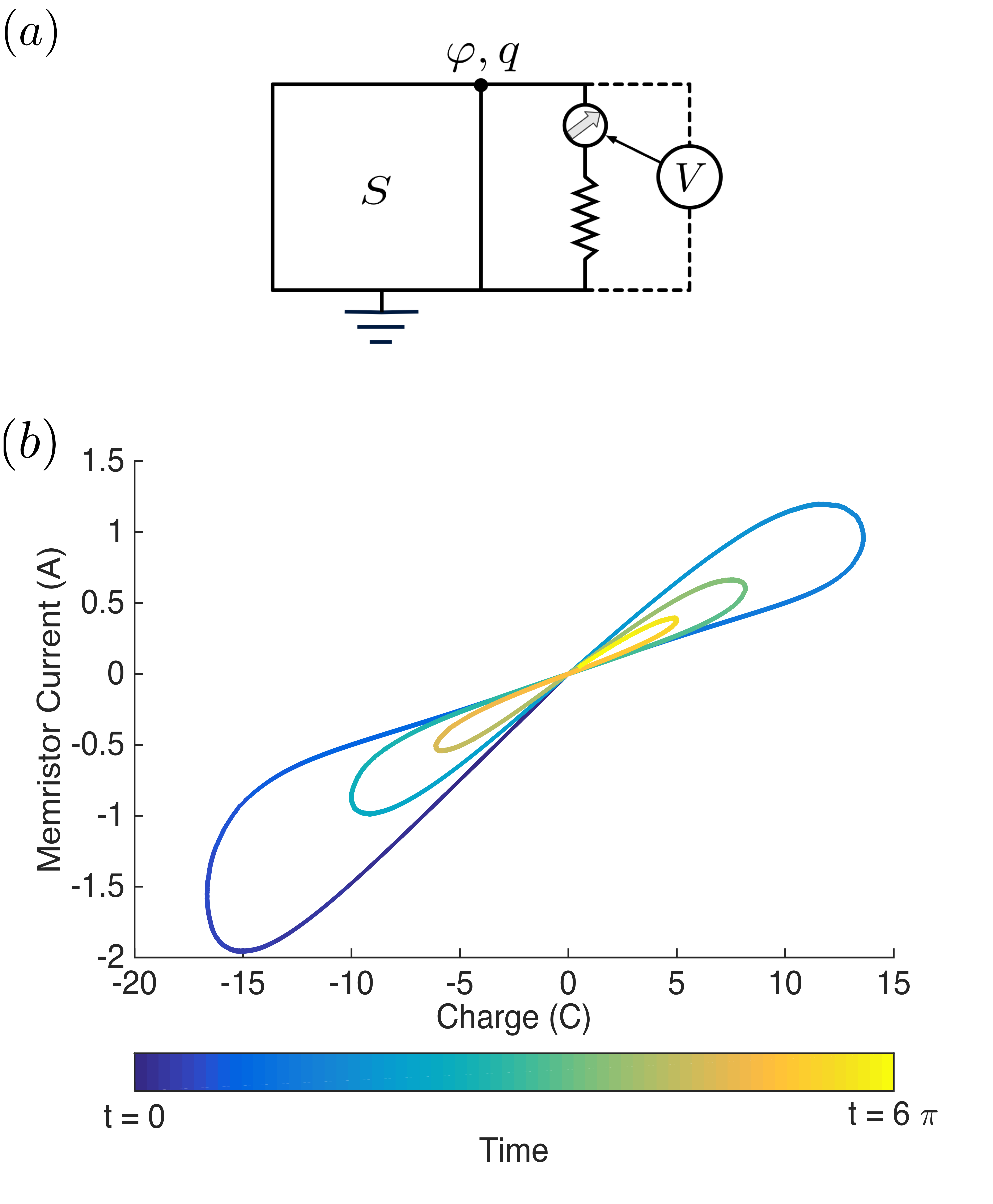}}
\caption{(a) A quantum memristor as a resistor coupled to a closed system with a voltage-based weak-measurement scheme. (b) Hysteresis loops  displayed by a quantum memristor coupled to an LC circuit with a system Hamiltonian $H_{S}= \frac{q^{2}}{2C}+\frac{\varphi^{2}}{2L}$, with $C=1$F and $L=1$H. Time is defined as the inverse of the frequency through three full cycles of the memristive device.}
\label{QMem}
\end{figure}

We now provide a brief introduction to circuit quantization. To do so, we first describe the quantization of the memristor. The quantum memristor was recently introduced~\cite{QMem}, and is described as a nonlinear element in a closed circuit with a weak-measurement scheme, which is used to update the resistance. This layout can be seen in Fig.~\ref{QMem}(a) as a closed system coupled to a resistor and a measurement apparatus, introducing a measurement-based update of the resistance depending on the system voltage.

The dynamics of the composite system can be studied by a master equation composed of Hamiltonian, continuous-weak-measurement, and dissipation parts:
\begin{equation}
d\rho=d\rho_H+d\rho_{m}+d\rho_{d}.
\end{equation}
The Hamiltonian part is given by the von Neumann equation:
\begin{equation}
d\rho_{H}=-\frac{i}{\hbar}[H_{S},\rho(t)] dt.
\end{equation}
The continuous-weak-measurement part reads
\begin{equation}
d\rho_{m}=-\frac{\tau}{q_0^2}[q, [q, \rho(t)]]dt+\sqrt{\frac{2\tau}{q_0^2}}(\{q, \rho(t)\}-2\langle q \rangle \rho(t))dW,
\end{equation}
where $[\cdot, \cdot]$ denotes a commutator and $\{\cdot, \cdot\}$ denotes an anticommutator. The expectation value of an observable is $\langle A \rangle=\tr(\rho A)$, $\tau$ is the projection frequency, $q_0$ is the uncertainty, the measurement strength is defined as $\kappa=\frac{\tau}{q_0^2}$, and $dW$ is the Wiener increment, which is related to the stochasticity associated with weak measurements.

The dissipation is described by a Caldeira-Leggett master equation:
\begin{equation}
d\rho_{d}=-\frac{i\gamma(\mu)}{\hbar}[\varphi, \{q, \rho(t)\}]dt-\frac{2C\lambda \gamma(\mu)}{\hbar}[\varphi, [\varphi, \rho(t)]]dt ,
\end{equation}
where $\lambda=k_{B}T/\hbar$ and $\gamma(\mu)$ is the relaxation rate. Solving these equations, we obtain the relation between memristive current and the charge shown in Fig.~\ref{QMem}(b), for an $LC$ circuit coupled to a memristor, with a Hamiltonian of the form $H_{S}= \frac{q^{2}}{2C}+\frac{\varphi^{2}}{2L}$. We observe the memristor displays the characteristic hysteresis curve when the current response is plotted versus the charge, which is related to the control voltage as $V=q/C$. In the case of a circuit with classical sources or a circuit coupled to an open element, there is no need to introduce the Wiener noise.

With a Hamiltonian formulation at hand, a description of electric circuits entails defining fluxes and charges, from which the voltage and the current can be obtained by time differentiation. In this case we use a node formulation, where node fluxes are the main variables and play the role of the spatial variable, with node charges being the conjugate variables. This formulation with node fluxes suffices to describe a circuit featuring linear capacitances and inductances.

\begin{figure}[t]
{\includegraphics[width=0.4 \textwidth]{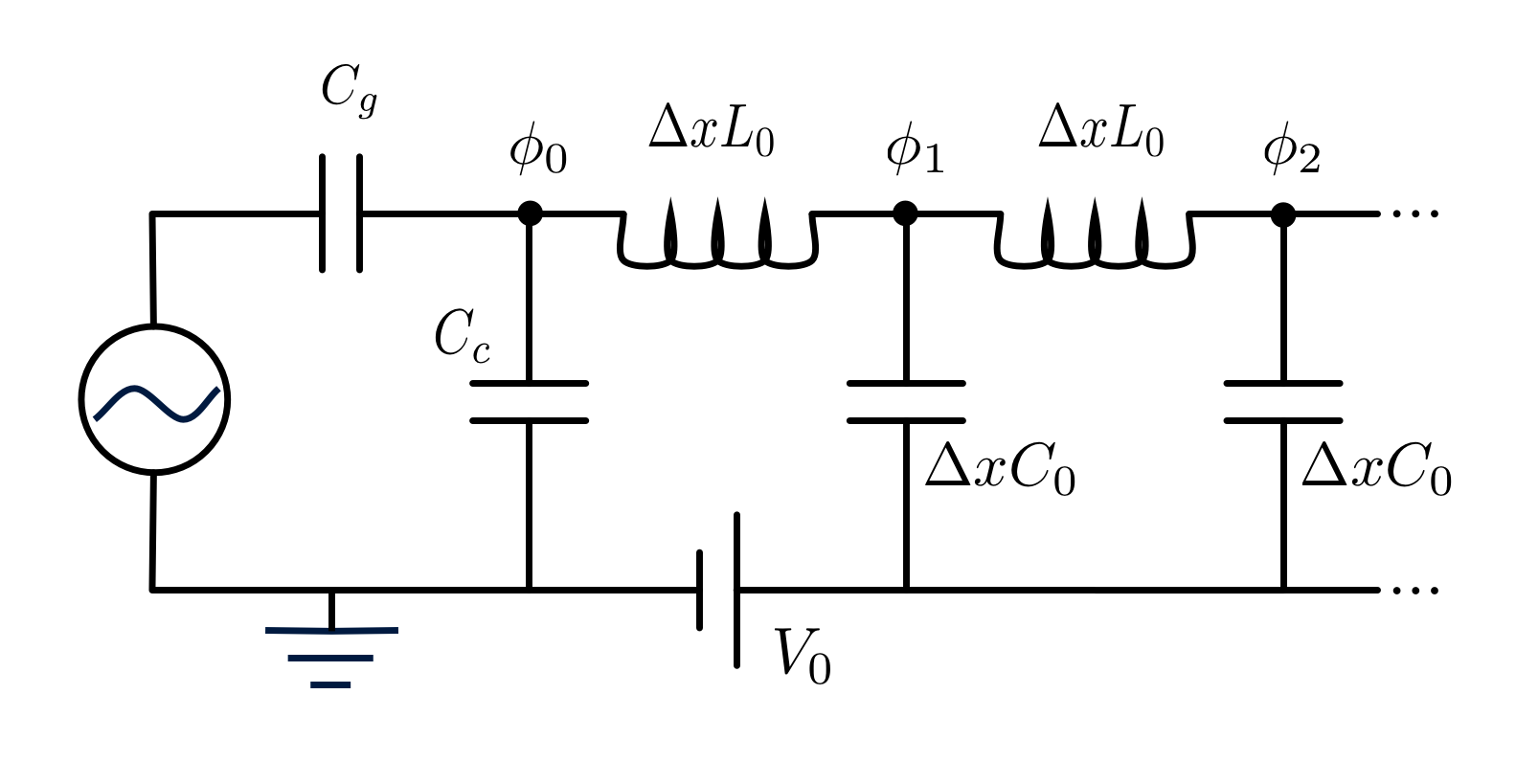}}
\caption{Hodgkin-Huxley circuit for a single ion channel with a classical ac source, $I(t)=I_{0}\sin(\Omega t)$, coupled to a semi-infinite transmission line. $C_{0}$ and $L_{0}$ are the capacitance and inductance corresponding to the transmission line, $C_{g}$ is the capacitance coupling the source to the circuit, $C_{c}$ accounts for the axon's membrane capacitance, and $V_{0}$ is the resting potential for the potassium-ion channel.}
\label{HHQ_tl}
\end{figure}

In a Lagrangian formalism, dissipative elements such as resistors can be treated by adding a dissipation function to the equations of motion of an effective Lagrangian~\cite{MemLagrangian}. However, the reversibility of Hamilton's equations, arising from a Hamiltonian formulation needed for a proper circuit quantization, conflicts with the irreversibility of dissipative terms. To describe the quantum memristor in a Lagrangian formulation suitable for canonical quantization, we assume linear dissipation and treat it as a linear dissipative element in the Caldeira-Leggett model~\cite{IntroQEC}. In this manner, we replace it by an infinite set of coupled $LC$ oscillators with a frequency-dependent impedance $Z(\omega)$; that is, a transmission line. We identify the impedance of the transmission line with the resistance of the memristor and, assuming that the time between consecutive updates is much larger than the memristor's relaxation time, the impedance can be updated (see Fig.~\ref{HHQ_tl}).

\section{Quantized Hodgkin-Huxley model}

\subsection{Classical input source}
We now study the quantization of the Hodgkin-Huxley circuit in Fig.~\ref{HHQ_tl}. For this, we write the Lagrangian describing this circuit:
\begin{equation}
\begin{split}
\mathcal{L}=&\frac{C_{g}}{2}(\dot{\phi}_{0}-\dot{\phi}_{s})^{2} + \frac{C_{c}}{2}\dot{\phi}^{2}_{0} - \frac{(\phi_{1}-\phi_{0})^{2}}{2\Delta x L_{0}} + \\
&+\sum^{\infty}_{i=1}\bigg[ \frac{\Delta x C_{0}}{2}(\dot{\phi}_{i}-V_{0})^{2}-\frac{(\phi_{i+1}-\phi_{i})^{2}}{2\Delta x L_{0}}\bigg].
\end{split}
\end{equation}
$C_{0}$ and $L_{0}$ are the characteristic capacitance and impedance per unit length of the transmission line, and the classical current source is defined as $I(t)=-C_{g}(\ddot{\phi}_{0}-\ddot{\phi}_{s})$. The motion of this circuit can be described by Euler-Lagrange equations, $\dv{}{t}\pdv{\mathcal{L}}{\dot{\phi}}=\pdv{\mathcal{L}}{\phi}$. Computing these equations for the intermediate node fluxes on the transmission line, $\phi_{i}$, we find 
\begin{equation}\label{waveq}
\ddot{\phi}_{i} =  \frac{1}{L_{0} C_{0}} \pdv[2]{\phi_{i}}{x}
\end{equation}
after taking the continuum limit, $\Delta x\rightarrow 0$. This is the wave equation for a flux field at position $x_{i}$ on the transmission line. The general solution of this equation can be written in terms of ingoing and outgoing waves, $\phi(x,t)=\phi_\text{in}(t + x/v) + \phi_\text{out}(t - x/v)$, with velocity $v=1/\sqrt{L_{0}C_{0}}$. This leads to the relations
\begin{equation}
\begin{split}
& \pdv{\phi(x,t)}{t} = \dot{\phi}_\text{in}(t+x/v) + \dot{\phi}_\text{out}(t-x/v),\\
& \pdv{\phi(x,t)}{x} = \frac{1}{v} (\dot{\phi}_\text{in}(t+x/v) - \dot{\phi}_\text{out}(t-x/v)),
\end{split}
\end{equation}
which allows us to obtain $\pdv{\phi_{0}(t)}{x} = \frac{1}{v} (2\dot{\phi}_\text{in}(t) - \dot{\phi}_{0}(t))$. The Euler-Lagrange equation for $\phi_{0}$ is
\begin{equation}
-I(t) + C_{c}\ddot{\phi}_{0} = \frac{1}{L_{0}}\pdv{\phi_{0}}{x}
\end{equation}
where we have taken the continuum limit. We can rewrite this as  
\begin{equation}\label{HHQ_phi_0}
-I(t) + C_{c}\ddot{\phi}_{0} + \frac{\dot{\phi}_{0}}{Z_{0}}= 2\frac{\dot{\phi}^\text{in}_{0}}{Z_{0}},
\end{equation}
where $Z_{0}=\sqrt{L_{0}/C_{0}}$ is the impedance of the transmission line, which is associated with the resistance of the memristor. $\phi_{0}(t)$ is our main variable because $\langle\dot{\phi}_{0}(t)\rangle$ will give us the circuit voltage for a given state of the transmission line. We have identified $\phi_{0}(t)$ with $\phi(x=0,t)$, the flux field inside the transmission line at $x=0$, and we want to quantize this flux.

Given that the wave equation inside the transmission line is satisfied, the flux field can be written in terms of ingoing and outgoing modes, fulfilling canonical commutation relations for a semi-infinite transmission line. This sets the starting point for the quantization of the field, which has been performed for infinite electrical networks~\cite{QNetwork}. To begin, we write the decomposition
\begin{equation}\label{chalmers}
\begin{split}
\phi(x,t)=&\sqrt{\frac{\hbar Z_{0}}{4\pi}}\int_{0}^{\infty}\frac{d\omega}{\sqrt{\omega}}(a_{in}(\omega) e^{i(k_{\omega}x - \omega t)}+\\
&+a_{out}(\omega) e^{-i(k_{\omega}x + \omega t)}+\text{H.c.}),
\end{split}
\end{equation}
where $k_{\omega}=|\omega|\sqrt{L_{0}C_{0}}$ is the wave vector. Here, $a_\text{in}(\omega)$ and  $a_\text{out}(\omega)$ can be promoted to quantum operators, and thus $\phi_{0}(t)$ is promoted to a quantum operator. By combining Eq.~(\ref{HHQ_phi_0}) with Eq.~(\ref{chalmers}) and writing the Fourier transform of the current, $I(t)=\int_{0}^{\infty} \frac{d\omega}{\sqrt{\omega}} (\mathcal{I}(\omega)e^{-i\omega t} + \mathcal{I}^{*}(\omega)e^{i\omega t})$, we can express the outgoing modes in terms of the ingoing ones:
\begin{equation}\label{reflection} 
a_\text{out}(\omega) = a_\text{in}(\omega) \frac{i - C_{c} \omega Z_{0}}{i + C_{c} \omega Z_{0}} - \frac{1}{\omega} \sqrt{\frac{4\pi Z_{0}}{\hbar}} \frac{\mathcal{I}(\omega)}{i + C_{c} \omega Z_{0}}.
\end{equation} 
$R(\omega) = \frac{i - C_{c} \omega Z_{0}}{i + C_{c} \omega Z_{0}}$ is the reflection coefficient, and we identify $s(\omega) = \frac{1}{\omega} \sqrt{\frac{4\pi Z_{0}}{\hbar}} \frac{\mathcal{I}(\omega)}{i + C_{c} \omega Z_{0}}$ as the source term, with $\mathcal{I}(\omega)=\frac{\sqrt{\omega}}{2\pi} \int^{\infty}_{-\infty} dt e^{i \omega t} I(t)$. Then the circuit voltage is
\begin{equation}
\begin{split}
& \langle\dot{\phi}_{0}(t)\rangle = -i \sqrt{\frac{\hbar Z_{0}}{4 \pi}} \int^{\infty}_{0} d\omega \sqrt{\omega} \\
& \bigg[\Big(\langle a_\text{in}(\omega)\rangle (1+R(\omega))-s(\omega) \Big)e^{-i \omega t} - \text{H.c.} \bigg]
\end{split}
\end{equation}
for a given state of the transmission line. By our choosing this state to be the vacuum, $\langle 0 | a_\text{in}(\omega) | 0 \rangle = \langle 0 | a^{\dagger}_\text{in}(\omega) | 0 \rangle = 0$, and for a classical input current $I(t)=I_{0}\sin(\Omega t)$, the voltage response of the system reads
\begin{equation}\label{HHQ_V}
\langle \dot{\phi}_{0} \rangle = I_{0}Z_{0} \frac{\sin(\Omega t) - C_{c} \Omega Z_{0} \cos(\Omega t)}{1 + (C_{c} \Omega Z_{0})^{2}}.
\end{equation}
This is what is obtained when the stationary solution of Eq.~(\ref{HHQ_phi_0}) with $\dot{\phi}^\text{in}_{0}=0$ is studied. However, we claim that $\phi_{0}$ is a valid quantum operator, and to demonstrate this we compute the second moment of the voltage. We do this for a vacuum state of the transmission line, meaning no excitations, and find
\begin{equation}
\begin{split}
\langle \dot{\phi}_{0}^{2} \rangle =&  \frac{\hbar Z_{0}}{\pi} \int^{\infty}_{0} d\omega \, \frac{\omega}{1+(C_{c} \omega Z_{0})^{2}} \,+\\
& + \Big[ Z_{0} I_{0} \frac{\sin(\Omega t) - C_{c} \Omega Z_{0} \cos(\Omega t)}{1 + (C_{c} \Omega Z_{0})^{2}} \Big]^{2}.
\end{split}
\end{equation}

The second term contains the voltage squared, which has a classical origin, at variance with the first term. The latter is related to the reflection of the modes in the circuit and is associated with the zero-point energy. It diverges as $\omega\rightarrow\infty$, which is a purely quantum-mechanical effect~\cite{IntroQEC}. We can eliminate the divergence in the voltage fluctuations by subtracting this quantity from the second moment of the voltage computed for the transmission line in a thermal state. This state is defined through Bose-Einstein statistics,
\begin{equation}
\begin{split}
& \langle a_{\text{in}}(\omega)a_{\text{in}}(\omega^{'})\rangle = \langle a_{\text{in}}^{\dagger}(\omega)a_{\text{in}}^{\dagger}(\omega^{'})\rangle = 0,\\
& \langle a_{\text{in}}^{\dagger}(\omega)a_{\text{in}}(\omega^{'})\rangle = \frac{1}{2}\Big[\coth\big(\frac{\beta\hbar\omega}{2}\big)-1\Big]\delta(\omega-\omega^{'}),\\
& \langle a_{\text{in}}(\omega)a_{\text{in}}^{\dagger}(\omega^{'})\rangle = \frac{1}{2}\Big[\coth\big(\frac{\beta\hbar\omega}{2}\big)+1\Big]\delta(\omega-\omega^{'}),
\end{split}
\end{equation}
for the number of bosonic modes. Here $\beta=1/k_{B}T$, where $k_{B}$ is the Boltzmann constant and $T$ is the temperature. We need to compute the following integral
\begin{equation}
\begin{split}
&\Delta \equiv \langle\dot{\phi}_{0}^{2}\rangle_{\text{thermal}} - \langle\dot{\phi}_{0}^{2}\rangle_{\text{vacuum}} =\\
& \frac{\hbar Z_{0}}{\pi}\int_{0}^{\infty}d\omega \frac{\omega}{1+(C_{c}\omega Z_{0})^{2}}\Big[\coth\big(\frac{\beta\hbar\omega}{2}\big)-1\Big],
\end{split}
\end{equation}
which by imposing $\hbar\beta\omega\gg 1$ reduces to
\begin{equation}
\frac{2\hbar Z_{0}}{\pi}\int_{0}^{\infty}d\omega \frac{\omega \, e^{-\beta\hbar\omega}}{1+(C_{c}\omega Z_{0})^{2}}
\end{equation}
Approximating the solution for large $\beta$ up to second order, we find 
\begin{equation}
\Delta = \frac{2 Z_{0}}{\hbar \pi \beta^{2}}.
\end{equation}
This gives the contribution of the quantum fluctuations of the circuit voltage without the zero-point energy. As expected, this quantity goes to zero as $T \rightarrow 0$. 

To observe the system memristive behavior, we introduce the update of the resistance of the memristor, considering the dependence of $Z_{0}$ on the circuit voltage. This approach is valid as long as the relaxation time of the set of $LC$ oscillators, which represents the instantaneous resistor, is much shorter than the timescale associated with the change in the resistance. This is equivalent to an adiabatic approximation~\cite{Adiabatic}, where we initially consider $Z_{0}$ as constant to obtain the value of the circuit voltage and it is consistently updated later.

\begin{figure*}[htp]
{\includegraphics[width=0.6 \textwidth]{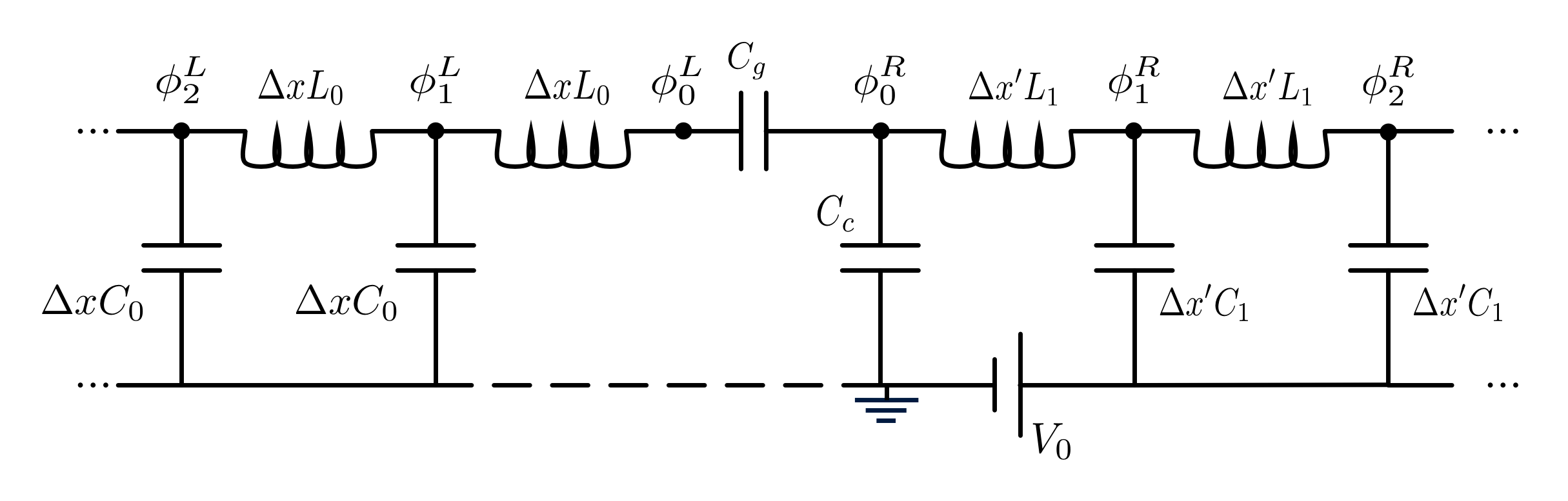}}
\caption{Hodgkin-Huxley circuit for a single ion channel coupled to a semi-infinite transmission line, introducing a quantized source on the left modeled by a second semi-infinite transmission line. $C_{0}$ and $L_{0}$ are the capacitance and inductance corresponding to the left transmission line and $C_{1}$ and $L_{1}$ are the capacitance and inductance corresponding to the right transmission line. $C_{g}$ is the capacitance coupling both transmission lines, $C_{c}$ accounts for the axon's membrane capacitance, and $V_{0}$ is the resting potential for the potassium-ion channel.}
\label{HHC_2TL}
\end{figure*}

Identifying the inverse of the impedance (known as admittance) with a conductance, we can use the potassium-channel conductance $g_{K}(t)=\bar{g}_{K}n(t)^{4}$ to update the impedance $Z(t) = Z_\text{min} n(t)^{-4}$, via Eq.~(\ref{mu1}), such that
\begin{equation}\label{Zupdate}
\dot{Z}(t) = -4 Z_\text{min} \bigg(\frac{Z(t)}{Z_\text{min}}\bigg)^{5/4}\alpha(V_{g}) + 4 Z(t) \Big(\alpha(V_{g}) + \beta(V_{g})\Big) .
\end{equation}
Thus, from now on, we identify the potassium-channel conductance with the inverse of the impedance of the transmission line. In this treatment, we assume that the voltage measurements used to update the impedance of the memristor do not perturb the system, considering measurements in this setup to be automatic and noninvasive. Given the dependence of the impedance on the circuit voltage, we solve Eq.~(\ref{Zupdate}) numerically to obtain the change in impedance of the memristor. Then we compute the potassium-channel conductance and the membrane voltage. This is equivalent, as mentioned above, to a strong adiabatic approximation, in which the impedance is considered constant in order to update the voltage of the circuit.

\subsection{Quantized input source}

To study the circuit response to quantum-state inputs, we replace the source with a second semi-infinite transmission line, thus introducing a collection of $LC$ circuits in which multiple frequencies can be excited. This circuit is depicted in Fig.~\ref{HHC_2TL}. In this scenario, the input current is represented by $\langle \dot{Q}^{L}_{0}\rangle$, where $Q^{L}_{0}=C_{g}(\dot{\phi}^{L}_{0}-\dot{\phi}^{R}_{0})$ for a given state that satisfies $\langle \dot{Q}^{L}_{0}\rangle=I_{0}\sin(\Omega t)$. The Lagrangian describing this system is
\begin{equation}
\begin{split}
\mathcal{L} =& \sum^{\infty}_{i=1}\bigg[ \frac{\Delta x C_{0}}{2}(\dot{\phi}^{L}_{i})^{2}-\frac{(\phi^{L}_{i}-\phi^{L}_{i+1})^{2}}{2\Delta x L_{0}}\bigg] -\frac{(\phi^{L}_{0} - \phi^{L}_{1})^{2}}{2\Delta x L_{0}}\\
& + \frac{C_{g}}{2}(\dot{\phi}^{R}_{0}-\dot{\phi}^{L}_{0})^{2}+\frac{C_{c}}{2}(\dot{\phi}^{R}_{0})^{2} - \frac{(\phi^{R}_{1}-\phi^{R}_{0})^{2}}{2\Delta x' L_{1}}\\
&+ \sum^{\infty}_{j=1}\bigg[ \frac{\Delta x' C_{1}}{2}(\dot{\phi}^{R}_{j}-V_{0})^{2}-\frac{(\phi^{R}_{j+1}-\phi^{R}_{j})^{2}}{2\Delta x' L_{1}}\bigg], 
\end{split}
\end{equation}

where the capacitance and inductance corresponding to the left transmission line are $C_{0}$ and $L_{0}$, and the capacitance and inductance corresponding to the right transmission line are $C_{1}$ and $L_{1}$. The Euler-Lagrange equations for $\phi^{L}_{0}$ and $\phi^{R}_{0}$ are
\begin{eqnarray}
 C_{g}(\ddot{\phi}^{L}_{0}-\ddot{\phi}^{R}_{0}) &=& \frac{2\dot{\phi}^{L}_{in}(t)}{Z_{0}} -  \frac{\dot{\phi}^{L}_{0}}{Z_{0}},\\
 C_{g}(\ddot{\phi}^{R}_{0}-\ddot{\phi}^{L}_{0}) + C_{c}\ddot{\phi}^{R}_{0} &=& \frac{2\dot{\phi}^{R}_{in}(t)}{Z_{1}} -  \frac{\dot{\phi}^{R}_{0}}{Z_{1}},
\end{eqnarray}
respectively. $Z_{0}$ is the impedance on the left transmission line and $Z_{1}$ is the impedance on the right transmission line. These equations can be written in this manner since the wave equation is satisfied inside each of the transmission lines for a flux field. We use the Euler-Lagrange equations to find expressions for the outgoing modes in terms of the ingoing ones:
\begin{equation}\label{R}
\begin{split}
&a^{L}_\text{out}(\omega) = a^{L}_\text{in}(\omega)R_{0}(\omega) + a^{R}_\text{in}(\omega) s(\omega),\\
&a^{R}_\text{out}(\omega) = a^{R}_\text{in}(\omega) R_{1}(\omega) + a^{L}_\text{in}(\omega) s(\omega).
\end{split}
\end{equation}
These modes have reflected and transmitted contributions on both sides of the circuit. The reflection coefficients are given by
\begin{equation}
\begin{split}
& R_{0}(\omega) =  \frac{1-i\omega(C_{g}+C_{c})Z_{1} + \omega C_{g}Z_{0}(i+\omega C_{c}Z_{1})}{1-i\omega(C_{g}+C_{c})Z_{1} - \omega C_{g}Z_{0}(i+\omega C_{c}Z_{1})},\\
& R_{1}(\omega) = \frac{1+i\omega(C_{g}+C_{c})Z_{1} - \omega C_{g}Z_{0}(i-\omega C_{c}Z_{1})}{1-i\omega(C_{g}+C_{c})Z_{1} - \omega C_{g}Z_{0}(i+\omega C_{c}Z_{1})},
\end{split}
\end{equation}
and the transmission coefficient is given by
\begin{equation}
 s(\omega) = \frac{-2i\omega C_{g}\sqrt{Z_{0} Z_{1}}}{1-i\omega(C_{g}+C_{c})Z_{1} - \omega C_{g}Z_{0}(i+\omega C_{c}Z_{1})}.
\end{equation}

Averaging over the vacuum state of the right transmission line, we write the voltage response of the circuit for a given state for the source
\begin{widetext}
\begin{equation}
\begin{split}
\langle \dot{\phi}^{R}_{0} \rangle =&  -C_{g} Z_{1} \sqrt{\frac{\hbar Z_{0}}{\pi}} \int^{\infty}_{0} d\omega \, \omega^{3/2} \, \bigg\{ \frac{ (1-\omega^{2}C_{g}C_{c}Z_{0}Z_{1})\big(\langle a^{L}_\text{in}(\omega)\rangle e^{-i\omega t} + \langle a^{L\dagger}_\text{in}(\omega)\rangle e^{i\omega t}\big)}{1+\omega^{2}\big(C_{g}^{2} Z_{0}^{2} + 2 C_{g}^{2}Z_{0}Z_{1} + ((C_{c}+C_{g})^{2}+\omega^{2}C_{g}^{2}C_{c}^{2}Z_{0}^{2})Z_{1}^{2}\big)}\\
& + \frac{i\omega\big((C_{c}+C_{g})Z_{1}+C_{g}Z_{0}\big)(\langle a^{L}_\text{in}(\omega)\rangle e^{-i\omega t} - \langle a^{L\dagger}_\text{in}(\omega)\rangle e^{i\omega t})}{1+\omega^{2}\big(C_{g}^{2} Z_{0}^{2} + 2 C_{g}^{2}Z_{0}Z_{1} + ((C_{c}+C_{g})^{2}+\omega^{2}C_{g}^{2}C_{c}^{2}Z_{0}^{2})Z_{1}^{2}\big)} \bigg\}.
\end{split}
\end{equation}
Computing the second moment of the voltage, we find
\begin{equation}
\begin{split}
\langle (\dot{\phi}^{R}_{0})^{2} \rangle =&  \frac{\hbar Z_{1}}{\pi} \int^{\infty}_{0} d\omega \, \frac{\omega(1+\omega^{2}C_{g}^{2}Z_{0}^{2})}{1+\omega^{2}\big(C_{g}^{2} Z_{0}^{2} + 2 C_{g}^{2}Z_{0}Z_{1} + ((C_{c}+C_{g})^{2}+\omega^{2}C_{g}^{2}C_{c}^{2}Z_{0}^{2})Z_{1}^{2}\big)} \\
& - \frac{\hbar Z_{1}}{4\pi} \int d\omega d\omega' \, \sqrt{\omega \omega'} \bigg\{\langle a^{L}_\text{in}(\omega) a^{L}_\text{in}(\omega')\rangle s(\omega)s(\omega') e^{-i(\omega + \omega') t} - \langle a^{L}_\text{in}(\omega) a^{L\dagger}_\text{in}(\omega')\rangle s(\omega)s^{*}(\omega') e^{-i(\omega - \omega') t} \\
& - \langle a^{L\dagger}_\text{in}(\omega) a^{L}_\text{in}(\omega')\rangle s^{*}(\omega)s(\omega') e^{i(\omega -\omega') t} + \langle a^{L\dagger}_\text{in}(\omega) a^{L\dagger}_\text{in}(\omega')\rangle s^{*}(\omega)s^{*}(\omega') e^{i(\omega + \omega') t} \bigg\},
\end{split}
\end{equation}
\end{widetext}
where the first term is again related to the reflection of the modes on the circuit, and it is purely quantum. The second term gives the frequency correlations of the modes on the left transmission line. 

The following step would be to find a quantum state of the source that gives the desired input current in the circuit, and to explore what quantum features can be used to enhance this model or to reveal interesting dynamics.

\begin{figure*}[htp]
{\includegraphics[width=0.95 \textwidth]{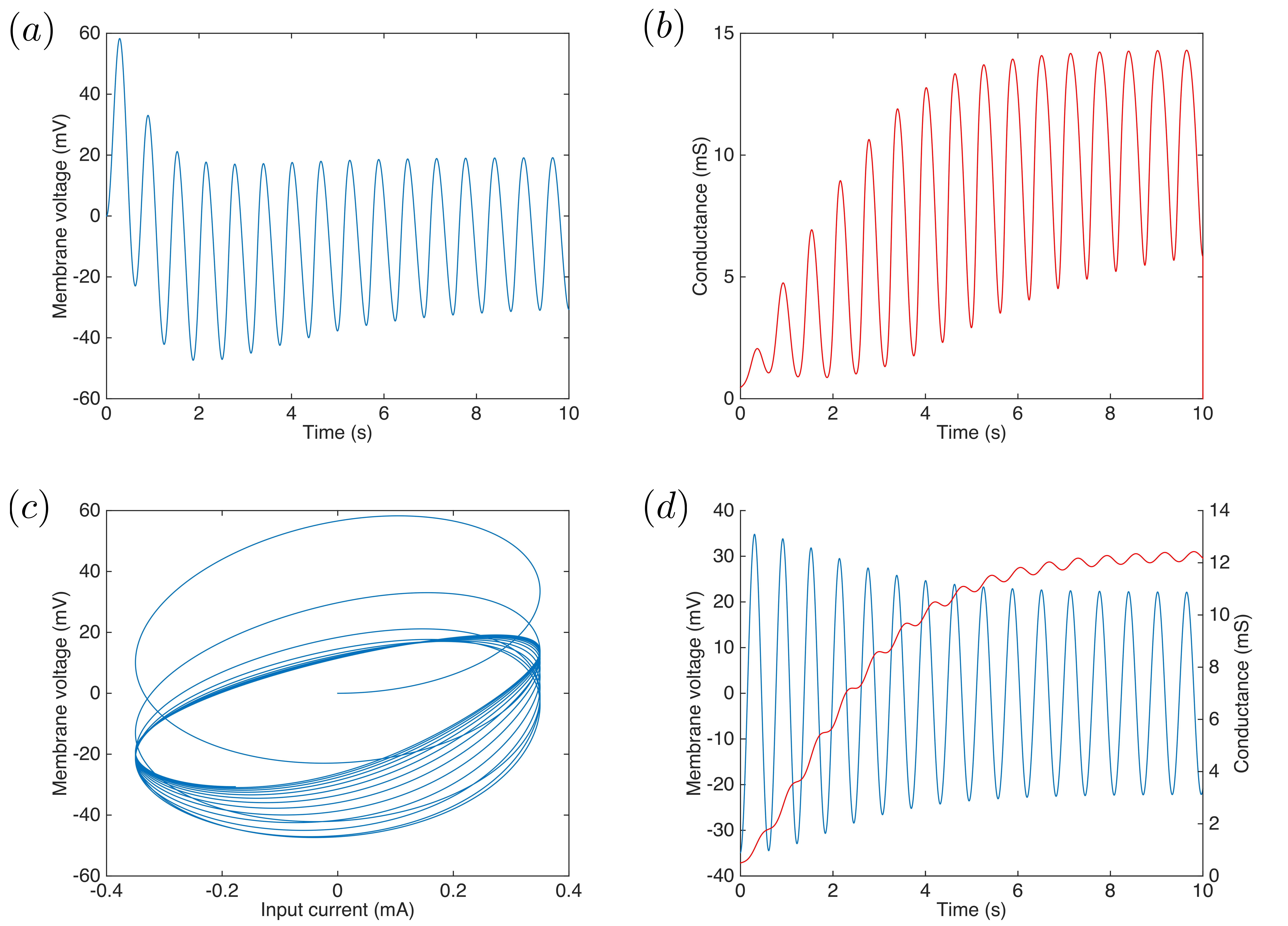}}
\caption{Classical Hodgkin-Huxley model for a single ion channel with a periodic input $I(t)=I_{0}\sin(\Omega t)$: (a) membrane voltage versus time; (b) potassium-channel conductance versus time; (c) membrane voltage versus input current; (d) membrane voltage (blue) and potassium-channel conductance (red) versus time with an adiabatic approximation.}
\label{HHC}
\end{figure*}

\section{Numerical Simulations}
In this section, we present the results for the membrane voltage and the potassium-channel conductance for the single-ion-channel Hodgkin-Huxley model. This is done by our solving Eq.~(\ref{Zupdate}) for the update of the memristor impedance using the results for the membrane voltage. We introduce the solutions for the membrane voltage and potassium-ion-channel conductance for the single-channel Hodgkin-Huxley model, which we use for comparison with the results for the quantized Hodgkin-Huxley model with a classical ac current source. In each case, we find that the potassium conductance reproduces qualitatively and approximately the S-shaped curve of Ref.~\cite{HHM}.

\subsection{Classical Hodgkin-Huxley model}

The single-channel Hodgkin-Huxley model with an ac current source is simulated by our solving the Hodgkin-Huxley equations; that is. Eq.~(\ref{current1}) together with Eq.~(\ref{mu1}). We plot in Fig.~\ref{HHC} the membrane voltage, the potassium-channel conductance, and the sodium-channel conductance, for a periodic input of the form $I(t)=I_{0}\sin(\Omega t)$. For the membrane voltage taken to be initially zero, we plot it versus time in Fig.~\ref{HHC}(a) as the blue curve; the red curve in Fig.~\ref{HHC}(b) corresponds to the potassium-channel conductance. 

A spike in the membrane voltage can be observed, with a subsequent decrease and adaptation to the input, leading to oscillations around the zero value. This is because we take the resting membrane voltage to be zero initially, and set $V_{\text{K}}=0$, where normally it is taken to be $V_{\text{K}}=-77$mV. We chose this according with the results for the response of the quantized model, in which this classical dc voltage source does not appear. This does not change the dynamical behavior; it just gives a displacement of the voltage. The profile of the voltage, aside from the oscillations caused by our choice of input current, accurately fits the plots depicted in Ref.~\cite{HHM}. The potassium conductance is reproduced with great accuracy according to Ref.~\cite{HHM}, featuring rising and adaptive behavior.

From comparison of the conductance as the red curve in Fig.~\ref{HHC}(b) with the ones obtained in Ref.~\cite{HHM} it can be appreciated that, with the introduction of an input signal, the conductance rises and adapts to this signal according to the depolarization of the membrane, with oscillations caused by our choice of $I(t)$. The $I-V$ curve in Fig.~\ref{HHC}(c) displays a hysteresis loop due to the periodic driving, which forms a limit cycle when the system saturates.

It is interesting to see that the spiking behavior of the membrane voltage can be reproduced in this simplified model featuring only potassium conductance. As the values for the coefficients $\alpha(V_{g})$ and $\beta(V_{g})$ were obtained through comparison with experimental results \cite{HHM}, the gate-opening probabilities for different ion channels may not be completely independent. Then, the mechanism of each ion channel cannot be isolated as the transmission of the nerve impulse is a balanced process involving (in this case two) different ion permeability changes.

When we solve the quantized Hodgkin-Huxley model, we use an adiabatic approximation. To fairly compare the classical and quantized models, we need to study the classical Hodgkin-Huxley model with an adiabatic approximation. The potassium conductance is given by $g_{K}=\bar{g}_{K}n^{4}(t)$. When solving Eqs.~(\ref{current1}) and~(\ref{mu1}), we consider $n(t)$ to be constant. Consequently, the straightforward solution of $V_{g}(t)$ with the choice of input current, $I(t)=I_{0}\sin(\Omega t)$, reads
\begin{equation}
V_{g}(t)=V_{\text{K}} + I_{0} \frac{g_{\text{K}}\sin(\Omega t)-\Omega C_{g}\cos(\Omega t)}{g_{\text{K}}^{2}+C_{g}^{2}\Omega^{2}} .
 \end{equation}
Here, we only consider the stationary solution, given that the transient one provides fast decay. Using this result, we solve $n(t)$ at any time step. We plot the results for the membrane voltage and the potassium-channel conductance as the blue curve and the red curve, respectively, in Fig.~\ref{HHC}(d). The results are exactly what we obtained with Eq.~(\ref{HHQ_V}), meaning that the voltage response of the system is classical, while the second moment of the voltage displays purely quantum terms.

\begin{figure*}[htp]
{\includegraphics[width=0.95 \textwidth]{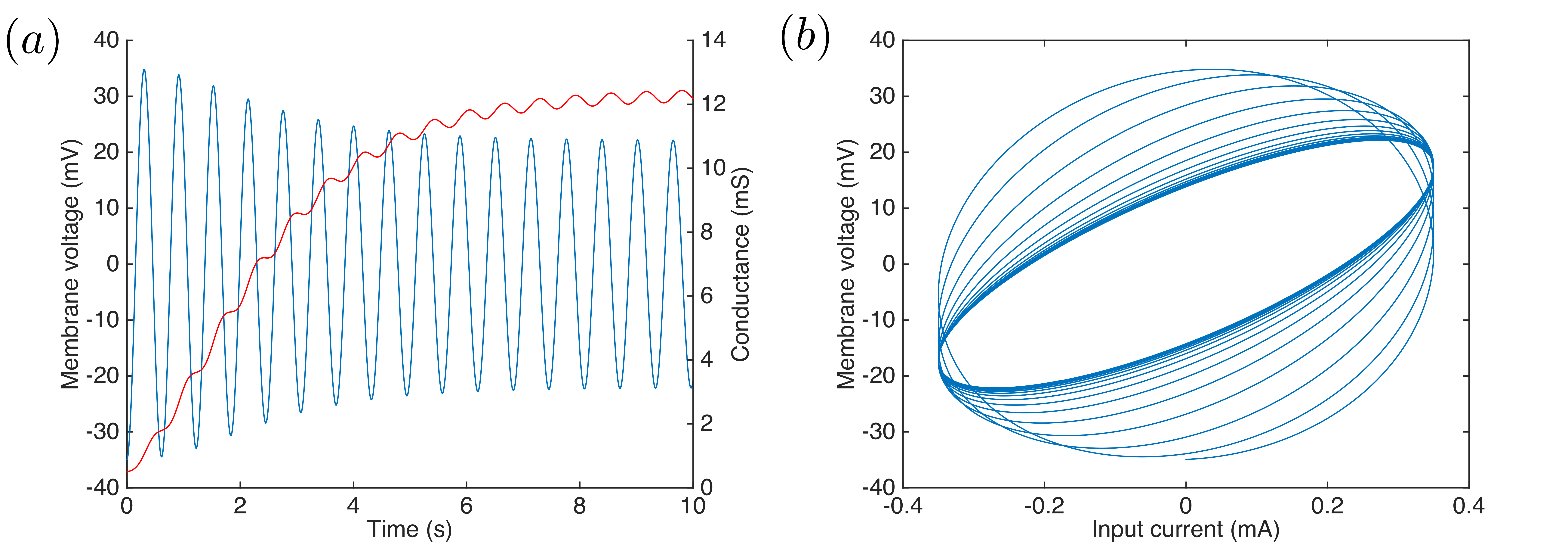}}
\caption{Quantum Hodgkin-Huxley model for a single ion channel with a classical periodic input $I(t)=I_{0}\sin(\Omega t)$ : (a) membrane voltage (blue) and potassium conductance (red) versus time; (b) membrane voltage versus input current.}
\label{HHQ}
\end{figure*}

\subsection{Quantized Hodgkin-Huxley model}

The simulation of the quantum Hodgkin-Huxley model uses a classical ac source. The solutions for the membrane voltage, the potassium-channel conductance, and the $I-V$ curve are presented below. By solving Eqs.~(\ref{HHQ_V}) and~(\ref{Zupdate}), we obtain the membrane voltage and the potassium conductance in the quantum model with a classical input $I(t)=I_{0}\sin(\Omega t)$. The membrane voltage is plotted against time as the blue curve in Fig.~\ref{HHQ}(a), where we observe a decrease in amplitude and a relaxation of the oscillations as it adapts to the input.

The conductance in Fig.~\ref{HHQ}(a) (red curve) does not feature a desired delay in its growth, but its saturation is clear, and it resembles the desired S-shaped curve displayed by the saturation of the potassium conductance in Ref.~\cite{HHM}. The membrane voltage and the potassium-channel conductance plotted in Fig.~\ref{HHQ}(a) are the same as in Fig.~\ref{HHC}(d), illustrating our statement that the response to a classical input source in the quantum regime is the same as in the classical regime with an adiabatic approximation. The membrane voltage is plotted versus the input current in Fig.~\ref{HHQ}(b), which features a memristive hysteresis loop. The system will have longer saturation times when the initial values are further away from the final value of the impedance. However, the system always relaxes into a limit cycle independent of the initial conditions.

The area of the hysteresis loop can give us a hint about the memory persistence in the system~\cite{CircuitQMem,OptQMem}, such that the larger the area, the greater the memory persistence. It would be interesting to test whether the introduction of a quantized Hodgkin-Huxley model, allowing the use of quantum-state inputs, is an improvement in the persistence of the memory. Particularly, entangled states are the desired states for this test. The information carried by quantum states can be related to classical information through Landauer's principle, classical dissipation being the link, where the area of the hysteresis loop intervenes. 

\begin{figure*}[htp]
{\includegraphics[width=0.95 \textwidth]{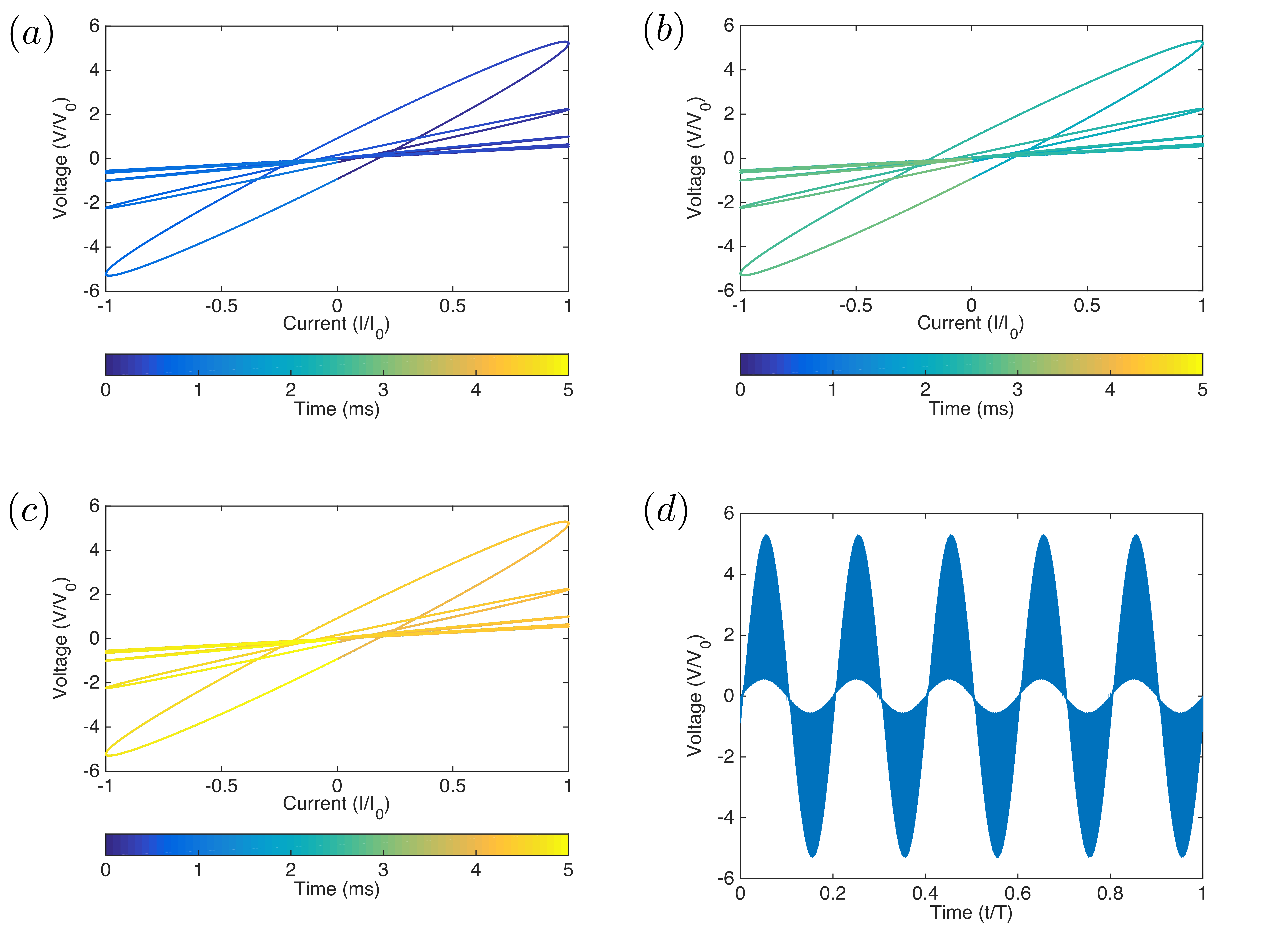}}
\caption{Numerical simulations of the quantized Hodgkin-Huxley model with a classical current source, introducing a quantum memristor engineered in a superconducting platform: (a) membrane voltage versus input current for one period of the driving current; (b) membrane voltage versus input current for three periods of the driving current; (c) membrane voltage versus input current for the duration of the potential spike; (d) membrane voltage versus time for the duration of the potential spike (5 ms). The membrane voltage and input current presented here are normalized to $V/V_{0}$ and $I/I_{0}$, respectively.}
\label{QHH_QMemSC}
\end{figure*}

\section{Feasibility of the implementation in superconducting circuits}
To complete our description of the Hodgkin-Huxley model in the quantum regime, we present a range in which the Hodgkin-Huxley circuit can be implemented in a superconducting platform. For that, we make use of the quantum memristor in superconducting circuits, as presented in Ref.~\cite{CircuitQMem}. This model consists of a superconducting loop interrupted by a dc SQUID with negligible loop inductance and threaded by a bias flux. In this circuit, the dc SQUID behaves as an effective flux-tunable Josephson junction controlled by the bias flux that threads the superconducting loop. In the latter, critical currents can be suppressed by applying to it a bias flux of half a flux quantum. Furthermore, the two junctions in the dc SQUID are chosen to be from different materials so that their conductances are different. In the limit of strong conductance asymmetry ($G_{1}\gg G_{2}$), tunneling of quasiparticles is favored ($\Delta_{1}\ll\Delta_{2}$), while critical currents can still be suppressed~\cite{MemSC}. This introduces nonlinearities in the circuit by means of a leakage current in the effective junction. 

A requirement for this description is that the bias flux is changed adiabatically to avoid the generation of quasiparticles, thus restricting the adiabatic parameter to $\text{max}\,\alpha_{\text{rs}} \approx 0.15$ for a sinusoidal driving of the bias flux. Furthermore, this device operates in the low-energy regime, $\hbar\omega_{10},\delta E \ll 2\Delta$, with a high-frequency requirement, $\hbar\omega_{10}\gg \delta E$, where, $\omega_{10}=\sqrt{2E_{C}E_{L}}/\hbar$ is the system transition frequency, $E_{C}$ and $E_{L}$ being the capacitive and inductive energies of the junction, respectively, and $\delta E$ is the characteristic energy of quasiparticles above the superconducting gap $\Delta \sim E_{L}$. The regime in which this implementation would function can be seen in Table~\ref{table1}.
\begin{table*}[]
\centering
\begin{tabular}{|>{\centering\arraybackslash} m{2.5 cm} |>{\centering\arraybackslash} m{2.5 cm} |>{\centering\arraybackslash} m{2.5 cm} |>{\centering\arraybackslash} m{2.5 cm} |>{\centering\arraybackslash} m{2.5 cm} |>{\centering\arraybackslash} m{2.5 cm} |}\hline
$E_{C}$ & $E_{L}$ & $\omega_{10}$ & $\Delta$ & $\delta E$ & $\alpha_{\text{rs}}$ \\ [2 ex] \hline
Capacitive energy & Inductive energy & Transition frequency & Superconducting gap energy & Quasiparticle energy & Adiabatic parameter \\ [2 ex] \hline
$2\pi\hbar$ GHz & $10^{3} \times 2\pi\hbar$ GHz & $44.72 \times 2\pi$ GHz & $\sim E_{L}$ & $\ll \hbar\omega_{10}$ & $\leq 0.15$ \\ [2 ex] \hline
\end{tabular}
\caption{Numerical values for the regime in which the quantum memristor is implemented in superconducting circuits (taken from Ref.~\cite{CircuitQMem}).}
\label{table1}
\end{table*}

The equations of a memristive system can be obtained  by means of the conductance of this device, which represents the connection between the quasiparticle current and the voltage across the total junction, after eliminating the critical currents in the dc SQUID. This way, we obtain a set of equations that mimic those of the classical memristor introduced in Ref.~\cite{MemSC}:  
\begin{eqnarray}
\nonumber && \langle \hat{I}_{\text{qp}} \rangle = G_{\text{qp}}[\langle\hat{\varphi}\rangle, \langle\hat{V}\rangle, t] \langle \hat{V} \rangle, \\
\nonumber && G_{\text{qp}}[\langle\hat{\varphi}\rangle, \langle\hat{V}\rangle, t] = G_{0} \sin^{2}\left[ \frac{\pi}{4} +\frac{1}{2}\sin\omega_{10}t \right], \\
\nonumber && \langle \hat{V} \rangle = V_{0} \cos\omega_{10}t,
\end{eqnarray}
where
\begin{eqnarray}
\nonumber && G_{0} = g_{0}^{2} e^{-g_{0}^{2}} \omega_{10}\frac{C_{d}}{2} \times 10^{-4} \approx 3.045 \times 10^{-9} \text{S}, \\
\nonumber && g_{0} = \left( \frac{E_{c}}{32 E_{L}}\right)^{1/4} \approx 7.476 \times 10^{-2},
\end{eqnarray}
and $\langle\hat{\varphi}\rangle$ is the internal variable of the quantum memristor. In this setting, the dynamics of the internal variable is suppressed in favor of the dynamics of the conductance. We then introduce this conductance as the inverse of the impedance in the equation for the membrane voltage in the Hodgkin-Huxley model:
\begin{equation}
\langle \dot{\phi}_{0}(t) \rangle = V_{g}(t) = I_{0}Z \, \frac{\sin(\Omega t) - C_{c}\Omega Z \cos(\Omega t)}{1+C_{c}^{2}\Omega^{2}Z^{2}}.
\end{equation}
In Table~\ref{table2} we give a regime in which the quantum memristor can be implemented in this architecture, and propose a complementary regime in which the behavior of the Hodgkin-Huxley circuit elements can be observed. At the same time, claims about the relaxation time of the memristor and the adiabatic approximation need to be fulfilled. 
\begin{table*}[]
\centering
\begin{tabular}{|>{\centering\arraybackslash} m{3cm} |>{\centering\arraybackslash} m{3cm} |>{\centering\arraybackslash} m{3cm} |>{\centering\arraybackslash} m{3cm} |>{\centering\arraybackslash} m{3cm} |}
\cline{1-5}
Architecture & $\omega$($2\pi$ Hz) & $T$(s) & $C$(F) & $G_{0}$ (S) \\ [2 ex] \cline{1-5}
Superconducting quantum memristor & $\omega_{10}\sim 4.5 \times 10^{10}$ & $T_{10}\sim 10^{-6}$ & $C_{d}\sim 5\times 10^{-13}$ & $3\times 10^{-9}$ \\ [2 ex] \cline{1-5}
Hodgkin-Huxley model & $\Omega\sim 10^{3}$ & $T_{\text{spike}}\sim 5\times 10^{-3}$ & $C_{c}\sim 10^{-13}$ & - \\ [2 ex] \cline{1-5}
\end{tabular}
\caption{Numerical values for the simulation of the Hodgkin-Huxley circuit with the equations of the quantum memristor engineered in superconducting circuits.}
\label{table2}
\end{table*}
These claims are fulfilled as the relaxation time of the memristor is given by $T_{10}$(on the order of microseconds), which is much smaller than the estimated duration of a potential spike, $T_{\text{spike}}$(approximately 5 ms), in the membrane of the axon. Also, the frequency of the driving is larger than that associated with the period of the spike. With these values, we are able to reproduce the hysteresis loop of the memristor when displaying the membrane voltage versus the input current, as we have $C_{c}\Omega/\bar{G} \sim 0.0328 \ll 1$ on average, which leads to
\begin{equation}
V(t) \sim \frac{I(t)}{G(t)} = \frac{I_{0}}{G_{0}} \, \frac{\sin(\Omega t)}{\sin^{2}\left[ \frac{\pi}{4} +\frac{1}{2}\sin(\omega_{10} t) \right]}
\end{equation}
In Fig.~\ref{QHH_QMemSC} we plot the membrane voltage versus the input current ($V/V_{0}$ vs $I/I_{0}$), which displays multiple hysteresis loops as the system relaxes. In this plot we can observe three different timescales, $T_{10}$ on the order of microseconds, $T_{\Omega}$ on the order of miliseconds, and $T_{\text{spike}}5\,ms$. We plot three different stages of the $I-V$ curve: (a) 1 ms, describing one period of the input-current source; (b) 3 ms, describing three periods of the input-current source; (c) 5 ms, describing 5 periods of the input current source, equivalent to the duration of the voltage spike in neurons. In Fig.~\ref{QHH_QMemSC} (d) we plot the membrane voltage ($V/V_{0}$) versus time for the duration of the voltage spike. We observe a modulation of the oscillations of the circuit voltage due to the different timescales in this implementation. 

The conductance in this case oscillates rapidly. As the internal dynamics of the quantum memristor are suppressed, we cannot see the rise of the potassium-channel conductance. A solution for this could be to study a model with two quantum memristors connected in parallel, but perhaps more effective would be to modify the superconducting qubit that produces memristive behavior. Adding another Josephson junction to the SQUID that interrupts the superconducting loop could add a flux that can be controlled. This way we can compensate for the suppression of the internal-variable dynamics.

\section{Conclusions \& Perspectives}
We study a simplified version of the Hodgkin-Huxley model with a single ion channel as a circuit featuring a capacitance, a voltage source, and a memristor, for a periodic input in both the classical regime and the quantum regime. The latter is achieved by our introducing the concept of a quantum memristor. Then we compare the membrane voltage, the potassium conductance, and the $I-V$ curve in both regimes.

This work shows that the behavior of this simplified version of the classical Hodgkin-Huxley model can be reproduced in the quantum regime. The voltage response of the circuit is found to be classical, but the second moment features a quantum-mechanical term given by the reflection of the modes in the circuit. The conductance is in good accordance with the experimental results reported in Ref.~\cite{HHM}, rising as an S-shaped curve. This is a result of a displacement from a resting value by an input source with an eventual adaptation, which is unaffected by intermediate and relaxation oscillations. This saturation or adaptation is identified with a learning process by the quantum memristor.

We also study the implementation of this circuit in current state-of-the-art quantum technologies by replacing the ion-channel quantum memristor by a superconducting quantum memristor. In this setup, we find a regime in which the properties of the superconducting quantum memristor can be observed, while the dynamics of the Hodgkin-Huxley circuit remain relevant.

A study of the two-ion channel Hodgkin-Huxley model in the quantum regime amounts to adding a second memristor corresponding to the conductance of the sodium channel~\cite{3QHH}. This would unveil new characteristics of the mechanism that rules the conduction of nerve impulses in neurons. Among other things, we would expect to see an initial spike in the sodium conductance, knowing that the mechanism of the sodium channel consists of a fast activation gate followed by inactivation. All this will require additional effort in the model: the study of two quantum memristors coupled in parallel. 

Another interesting line to follow is to study the effects of quantum-state inputs on the system, where memory effects are revealed by the area of the hysteresis loop. However, memory effects are more relevant when displayed in connected neuron networks. Studying, for example, the output of a string of neurons with an entangled-state input would imply yet another novel discovery: the dynamics of two circuits connected in series involving quantum memristors. Recent work describing the dynamics of memristor circuits coupled in series and in parallel~\cite{CoupledMem, CompositeMem, TheoremMem} can answer these two questions when taken to a quantum regime. This would set excellent starting points for any advances in neuromorphic quantum computing and quantum neural networks, with direct applications for quantum machine learning. 

\acknowledgements
The authors acknowledge support from Spanish Government PGC2018-095113-B-I00 (MCIU/AEI/FEDER, UE) and Ram\'on y Cajal Grant RYC-2017-22482, and Basque Government IT986-16, as well as Shanghai Municipal Science and Technology Commission (18010500400 and 18ZR1415500), and the Shanghai Program for Eastern Scholar. The authors also acknowledge support from the projects QMiCS (820505) and OpenSuperQ (820363) of the EU Flagship on Quantum Technologies, as well as EU FET Open Grant Quromorphic (828826). This material is also based upon work supported by the U.S. Department of Energy, Office of Science, Office of Advance Scientific Computing Research (ASCR), under field work proposal number ERKJ333.

\end{document}